\documentclass[prl,aps,twocolumn,pdftex]{revtex4-1}
\usepackage{graphicx}
\usepackage{amsmath}
\usepackage{graphicx}

\begin{document}
\title{Identifying the Charge Carriers of the Quark-Gluon Plasma}
\author{Scott Pratt}
\affiliation{Department of Physics and Astronomy and National Superconducting Cyclotron Laboratory,
Michigan State University\\
East Lansing, Michigan 48824}
\date{\today}
\begin{abstract}
Charge correlations in lattice gauge calculations suggest that up, down and strange charges move independently in the QGP (quark-gluon plasma), and that the density of such charges is similar to what is expected from simple thermal arguments. Here, we show how specific elements of the charge-charge correlation matrix in the QGP survive hadronization and become manifest in final-state charge-charge correlation measurements.
\end{abstract}

\maketitle

Identifying partonic degrees of freedom in the QGP (quark-gluon plasma) is a principal goal for the relativistic heavy ion programs at RHIC (Relativistic Heavy Ion Collider) and at the LHC (Large Hadron Collider). From the strong collective flow observed at RHIC, it is clear that the QGP interacts strongly and behaves like a liquid \cite{rhicwhitepapers}. The low viscosity suggests that collisional widths of partons are similar in magnitude to their energy, so quarks and gluons are not particularly good quasi-particles. Nonetheless, they can still serve as the fundamental carriers of charge. Lattice gauge calculations confirm the picture that up, down and strange charges are transported as single charges, in contrast to the hadronic world where they are transported in either charge-anticharge pairs (mesons) or in groups of three charges (baryons) \cite{Borsanyi:2011sw,Bazavov:2012jq}. Insight from lattice calculations comes from charge-charge correlations,
\begin{equation}
\chi_{ab}\equiv{\langle Q_aQ_b\rangle}/{V},
\end{equation}
where within the volume $V$, $Q_a$ is the net number of up, down or strange quarks. Here, we consider systems with zero net charge, otherwise one would subtract terms $\langle Q_a\rangle\langle Q_b\rangle$. If each charge carrier had one unit charge and moved independently, as in a parton gas, the off-diagonal elements would be zero and the correlation would be
\begin{equation}
\label{eq:chi_gas}
\chi^{\rm(parton~gas)}_{ab}=\delta_{ab}\left(n_a+n_{\bar{a}}\right).
\end{equation}
For example, $n_u$ is the density of up quarks and $n_{\bar{u}}$ is the density of anti-up quarks. The non-zero correlation comes from charges being correlated with themselves. This contrasts with a hadronic gas,
\begin{equation}
\label{eq:hadron_gas}
\chi^{\rm(hadron~gas)}_{ab}=\sum_\alpha n_{\alpha}q_{\alpha,a}q_{\alpha,b}~,
\end{equation}
where $q_{\alpha,a}$ is the charge of type $a$ on the hadronic species $\alpha$. Mesons tend to make the off-diagonal elements negative while baryons have the opposite effect. Within the last few years lattice calculations have provided $\chi_{ab}$ as a function of temperature, and shown the transition from a hadron gas to a QGP \cite{Borsanyi:2011sw,Bazavov:2012jq}. For temperatures above the critical region, the off-diagonal elements largely disappear and the magnitude of the diagonal elements approach 90\% of the expected value for a parton gas. In this letter we show how $\chi_{ab}$ from the QGP phase becomes manifest in experimental measurements.

Local charge conservation precludes correlations from lattice calculations being directly compared to those in experiment. Strictly speaking, the net charge in the complete volume is fixed and does not fluctuate, whereas in lattice calculations one assumes interaction with a particle sink, i.e., the grand canonical ensemble. However, charge correlations binned as a function of relative coordinate can be analyzed, and depend on $\chi_{ab}$ as calculated in a grand canonical ensemble. For the partonic stage in a heavy-ion collision, we assume the following form for the binned correlation,
\begin{eqnarray}
\label{eq:gabgas}
g_{ab}(\eta,\tau)&\equiv&\langle\rho_a(0)\rho_b(\eta)\rangle,\\
\nonumber
&\approx&\left(n_{a}+n_{\bar{a}}\right)\delta_{ab}
\left(\delta(\eta)-\frac{e^{-\eta^2/2\sigma_{\rm QGP}^2}}{(2\pi\sigma^2_{\rm QGP})^{1/2}}\right).
\end{eqnarray}
Since quarks are relatively light ($m<T$), the number of quarks is roughly fixed between the initial thermalization of the QGP and hadronization. Balancing charges then diffuse away from one another with a separation described by $\sigma_{\rm QGP}$. The delta function represents the correlation of a charge with itself, and if the balancing charges were allowed to spread over an infinite volume it would be the sole term and one would reproduce the parton gas value for $\chi_{ab}$ in Eq. (\ref{eq:chi_gas}). The relative coordinate along the longitudinal axis is the spatial rapidity, $\eta=\tanh^{-1}(z/t)$, and is known as the spatial rapidity, and is chosen instead of the coordinate $z$ to better account for the longitudinal expansion, which for a boost invariant  system is $v_z=z/t$. In a boost invariant system correlations depend only on the relative spatial rapidity $\eta$. The width $\sigma_{QGP}(\tau)$ is then frozen in the limit of zero diffusion. From this point forward, charge and number densities are per unit $\eta$, and from global charge conservation, $\int d\eta ~g_{ab}(\eta)=0$. For an adiabatic expansion the entropy and quark numbers per unit $\eta$ are approximately fixed, and in the gas limit $\chi_{ab}$ would stay constant during the evolution of the $QGP$. The proper time $\tau=\sqrt{t^2-z^2}$ is the time measured by an observer moving with the fluid from the origin $z=t=0$.

More generally, the strength of the Gaussian in the expression for $g_{ab}(\eta)$ in Eq. (\ref{eq:gabgas}) is $\chi^{\rm(QGP)}_{ab}$, which allows one to compare to lattice calculations. Unfortunately, measurements are not made before hadronization, but only afterward, when particles are on their final-state trajectories. To connect to experiment one must overcome three challenges:
\begin{enumerate} \itemsep 0pt
\item[a)] A second wave of quark-antiquark pairs are created or destroyed at hadronization.
\item[b)] Hadrons are measured -- not quarks.
\item[c)] One can measure only momenta, not spatial rapidity. Furthermore, strangeness violating decays affect the conservation rules. 
\end{enumerate}

To overcome (a) one uses the fact that $g_{ab}(\eta)$ cannot change suddenly except at $\eta\sim 0$ due to local charge conservation. Additionally, the correlation afterward must integrate to zero. Assuming hadronization is relatively sudden, the correlation after hadronization should have the form,
\begin{eqnarray}
\label{eq:gab_had}
\nonumber
g_{ab}^{\prime{(\rm HAD)}}(\eta)&=&-\left(\chi_{ab}^{\rm (HAD)}-\chi_{ab}^{\rm(QGP)}\right)
\frac{e^{-\eta^2/2\sigma^2_{\rm HAD}}}{(2\pi\sigma^2_{\rm HAD})^{1/2}}\\
&&-\chi_{ab}^{\rm(QGP)}\frac{e^{-\eta^2/2\sigma_{\rm QGP}^2}}{(2\pi\sigma^2_{\rm QGP})^{1/2}}.
\end{eqnarray}
The prime on $g^{\prime\rm(HAD)}$ denotes that the correlation of a particle with itself is subtracted out. The width $\sigma_{\rm HAD}$ represents the separation of those balancing charges created or destroyed at, or just after, hadronization and should be significantly smaller than $\sigma_{\rm QGP}$. The quantity $\chi^{\rm(HAD)}_{ab}$ is defined in Eq. (\ref{eq:hadron_gas}), and since the yields $n_\alpha$ are measured, $\chi_{ab}^{\rm (HAD)}$ is a known quantity. The form for $g'_{ab}$ in Eq. (\ref{eq:gab_had}) comes from freezing the long-range part, while constraining $g_{ab}$ to integrate to zero, or equivalently, enforcing $g'_{ab}$ to integrate to $-\chi^{\rm(HAD)}_{ab}$. Thus, $g'_{ab}(\eta)$ depends on: $\chi_{ab}^{\rm(QGP)}$, which can be taken from Eq. (\ref{eq:chi_gas}) or even directly from lattice calculations, $\chi_{ab}^{\rm(HAD)}$, which can be extracted from measured hadronic yields, and the width $\sigma_{\rm QGP}$. 

For (b) it was shown in \cite{Pratt:2011bc} how to use $g^{\prime\rm(HAD)}$ to generate correlations between hadronic species,
\begin{equation}
\label{eq:Gdef}
G_{\alpha\beta}(\eta)\equiv\langle[n_\alpha(0)-n_{\bar{\alpha}}(0)][n_\beta(\eta)-n_{\bar{\beta}}(\eta)]\rangle,
\end{equation}
where the Greek subscripts $\alpha$ and $\beta$ refer to specific hadronic species. 
To derive $G_{\alpha\beta}(\eta)$ from $g'_{ab}(\eta)$ one assumes that correlations are distributed thermally amongst species, i.e.,
\begin{equation}
\label{eq:mudef}
\langle n_\alpha(0)n_\beta(\eta)\rangle=\langle n_\alpha(0)\rangle\langle n_\beta(\eta)\rangle
e^{\mu_{ab}(\eta)q_{\alpha,a}q_{\beta,b}},
\end{equation}
where $\mu_{ab}$ is a Lagrange multiplier responsible for constraining $g'_{ab}$. Because $g'_{ab}$ can be determined from $G_{\alpha\beta}$,
\begin{equation}
g'_{ab}(\eta)=(1/4)\sum_\alpha G_{\alpha\beta}(\eta)q_{\alpha,a}q_{\beta,b}~,
\end{equation}
one can write $g'$ in terms of $G$, then use Eq. (\ref{eq:Gdef}) to write $G$ in terms of correlations of densities, which using Eq. (\ref{eq:mudef}) above to ultimately express $g'(\eta)$ in terms of $\mu(\eta)$ and measured densities $\langle n_\alpha\rangle$. In \cite{Pratt:2011bc} it was shown how to invert the expression and find $\mu$ in terms of $g'$ assuming $\mu$ is small. This led to an expression for $G$ in terms of $g'$,
\begin{eqnarray}
\label{eq:finalresult}
\nonumber
G_{\alpha\beta}(\eta)&=&4\sum_{abcd}\langle n_{\alpha}(0)\rangle
q_{\alpha,a}\chi^{{\rm(HAD)}(-1)}_{ac}(0)g^{\prime{\rm(HAD)}}_{cd}(\eta)\\
&&\hspace*{18pt}\cdot\chi^{{\rm(HAD)}(-1)}_{db}(\eta)
q_{\beta,b}\langle n_{\beta}(\eta)\rangle.
\end{eqnarray}
Inserting the exprssion for $g'$ in Eq. (\ref{eq:gab_had}) into Eq. (\ref{eq:finalresult}) one sees that $G_{\alpha\beta}(\eta)$ between any two hadronic species is determined by $\chi_{ab}^{\rm(QGP)}$, the measured yields of hadrons, and two unknown parameters $\sigma_{\rm QGP}$ and $\sigma_{\rm HAD}$. One doesn't expect significant sensitivity to $\sigma_{\rm HAD}$ since it should be smaller than the thermal spread, so effectively the unknowns in calculating $G_{\alpha\beta}(\eta)$ are $\chi_{ab}^{\rm(QGP)}$ and $\sigma_{\rm QGP}$.

Finally, to overcome (c), the correlations in spatial rapidity must be mapped to those in regular rapidity $y=\tanh^{-1}(v_z)$. Thermal smearing in the mapping of $G_{\alpha\beta}(\eta)\rightarrow G_{\alpha\beta}(y)$ and the decays of strange hadrons are taken into account with a blast-wave model \cite{Pratt:2011bc}. The blast-wave model was based on a gaussian distribution of transverse flow velocities (with a variance of 0.6$c$) and a fixed kinetic breakup temperature (120 MeV). Parameters were set to match mean transverse momenta for pions and protons measured by the STAR collaboration at RHIC \cite{Adams:2003xp}. After generating correlated pairs with the weights described by $G_{\alpha\beta}$, the decays of hyperons and neutral kaons were simulated. The final products along with their weights were binned to generate $G_{\alpha\beta}(y)$. Details of the method can be found in \cite{Pratt:2011bc}, though the blast-wave model used in this study is slightly different since it uses a Gaussian profile. The fit to baryon yields was accomplished by assuming a chemically equilibrated sample with a temperature of 165 MeV, combined with a suppression factor for baryons of two thirds to roughly match proton yields measured by PHENIX \cite{Adler:2003cb}. The default calculation assumes that the density of all quarks during the QGP is 0.85 times the final-state density of hadrons, and that the density of strange quarks is 92\% that of up or down quarks. This last assumption effectively sets the default $\chi^{\rm(QGP)}_{ab}$. Aside from a factor of the yields, $G_{\alpha\beta}(y)$ is equivalent to charge-balance functions which have been studied in depth \cite{Bass:2000az,Cheng:2004zy,Schlichting:2010qia,Bozek:2004dt} and measured for a few species \cite{Aggarwal:2010ya,Adams:2003kg,Westfall:2004cq}.

\begin{figure}
\centerline{\includegraphics[width=0.4\textwidth]{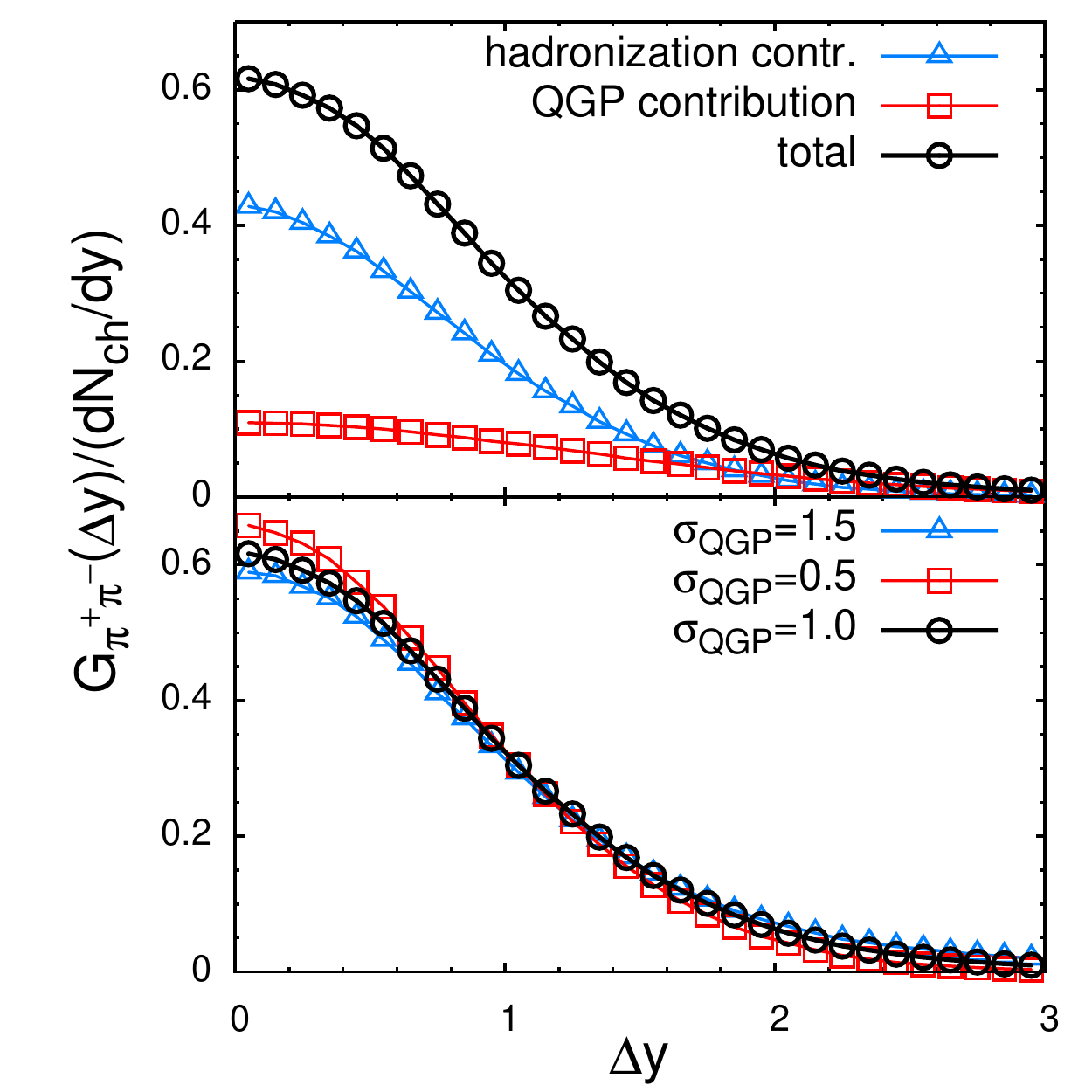}}
\caption{\label{fig:pipi}
The correlation for $\pi^+\pi^-$ scaled by multiplicity and separated by the QGP and hadronization contributions in the upper panel. The tail of the correlation function is dominated by the QGP portion, and is more pronounced for larger $\sigma_{\rm QGP}$ as is seen in the lower panel.}
\end{figure}
The goal of this letter is to illustrate how $G_{\alpha\beta}(y)$ is sensitive to specific elements of $\chi_{ab}^{\rm(QGP)}$ and $\sigma_{\rm QGP}$. Figure \ref{fig:pipi} shows how the width of the correlation $G_{\pi^+\pi^-}$ depends on $\sigma_{\rm(QGP)}$. For pions, roughly two thirds of the correlation comes from the hadronization component (with a characteristic spread of $\sigma_{\rm HAD}$, while approximately one third comes from the QGP component. This is expected given that well over half of the up and down quarks should be created at hadronization \cite{Bass:2000az}. The fraction can be understood through entropy arguments, which suggest that the number of quarks in the QGP and final-state hadrons are roughly equivalent, but since hadrons have multiple quarks, the number must more than double at hadronization. In Fig. \ref{fig:pipi} $\sigma_{\rm HAD}$ was set to 0.2, while results for three different values of $\sigma_{\rm(QGP)}$ are shown, the default value of 1.0, along with 0.5 and 1.5. Currently, there are no good experimental handles on $\sigma_{\rm QGP}$, so an analysis like this is required. One can make estimates of the diffusive width with crude diffusion models \cite{Bass:2000az}. Such estimates are $\gtrsim 0.5$, but may understate the case since the balancing charges created at the formation of the QGP may also be separated due to the tunneling associated with the breaking of color flux tubes. The tail of the correlation in Fig. \ref{fig:pipi} is clearly sensitive to $\sigma_{\rm QGP}$, but will be difficult to distinguish unless one has a detector with a wide acceptance in rapidity. The STAR detector at RHIC and the ALICE detector at the LHC cover ranges of relative rapidity close to 2 units, which may make the determination of $\sigma_{\rm QGP}$ difficult. Although the CMS and ATLAS detectors do not have particle ID for a large rapidity range, they can identify charges, and cover well over three units of rapidity. This should be sufficient for determining $\sigma_{\rm QGP}$ at LHC energies.

\begin{figure}
\centerline{\includegraphics[width=0.4\textwidth]{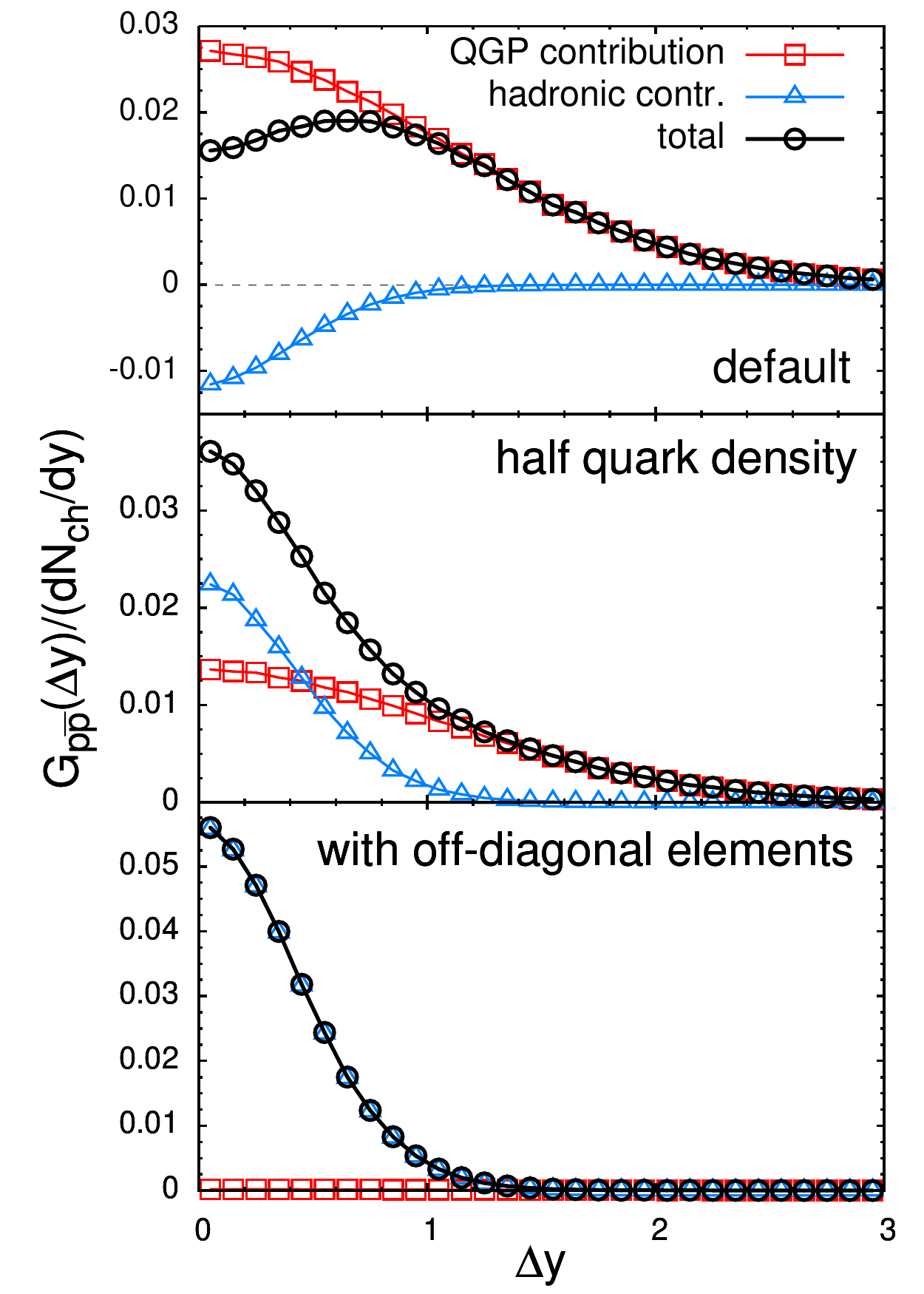}}
\caption{\label{fig:ppbar}
Correlations involving other hadronic species are more effective at separating the QGP and hadronization components because the two components might contribute with opposite signs. Such is the case for the $p\bar{p}$ and $pK^-$ correlations shown in the upper panels of Figs. \ref{fig:ppbar} and \ref{fig:pKminus}, respectively. The dip in the $p\bar{p}$ correlation and the switch of signs in the $pK^-$ correlation only exist if there are competing correlations of opposite sign. Proton-antiproton correlations test the idea that charge production comes in two separate waves, because the hadronization and QGP contributions have opposite signs as seen in the upper panel. The middle panel illustrates the sensitivity to reducing the diagonal elements of $\chi^{\rm(QGP)}_{ab}$ by a factor of two, and the effect of adding large off-diagonal components as in Eq. (\ref{eq:offdiag}) is displayed in the lower panel.
}
\end{figure}
\begin{figure}
\centerline{\includegraphics[width=0.4\textwidth]{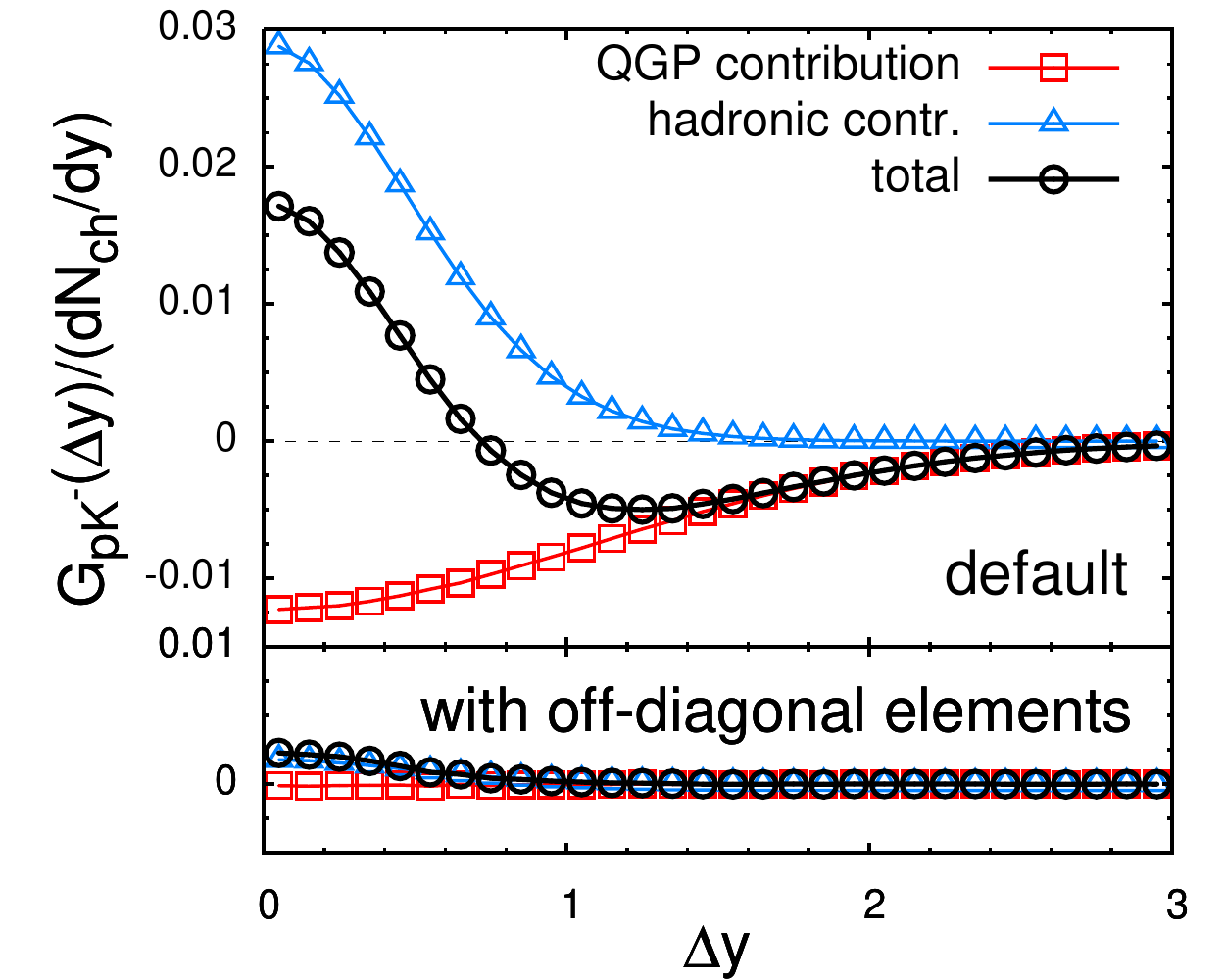}}
\caption{\label{fig:pKminus}
The hadronization and QGP components of the $pK^-$ correlations, shown in the upper panel, have opposite signs, which results in a correlation that changes sign. This would represent dramatic evidence of the two-wave nature of charge production. The $pK^-$ correlation is sensitive to the off-diagonal elements of $\chi^{\rm(QGP)}$, as can be seen by comparing the upper panel where there are no such elements to the lower panel which uses the elements described by Eq. (\ref{eq:offdiag}).
}
\end{figure}
\begin{figure}
\centerline{\includegraphics[width=0.4\textwidth]{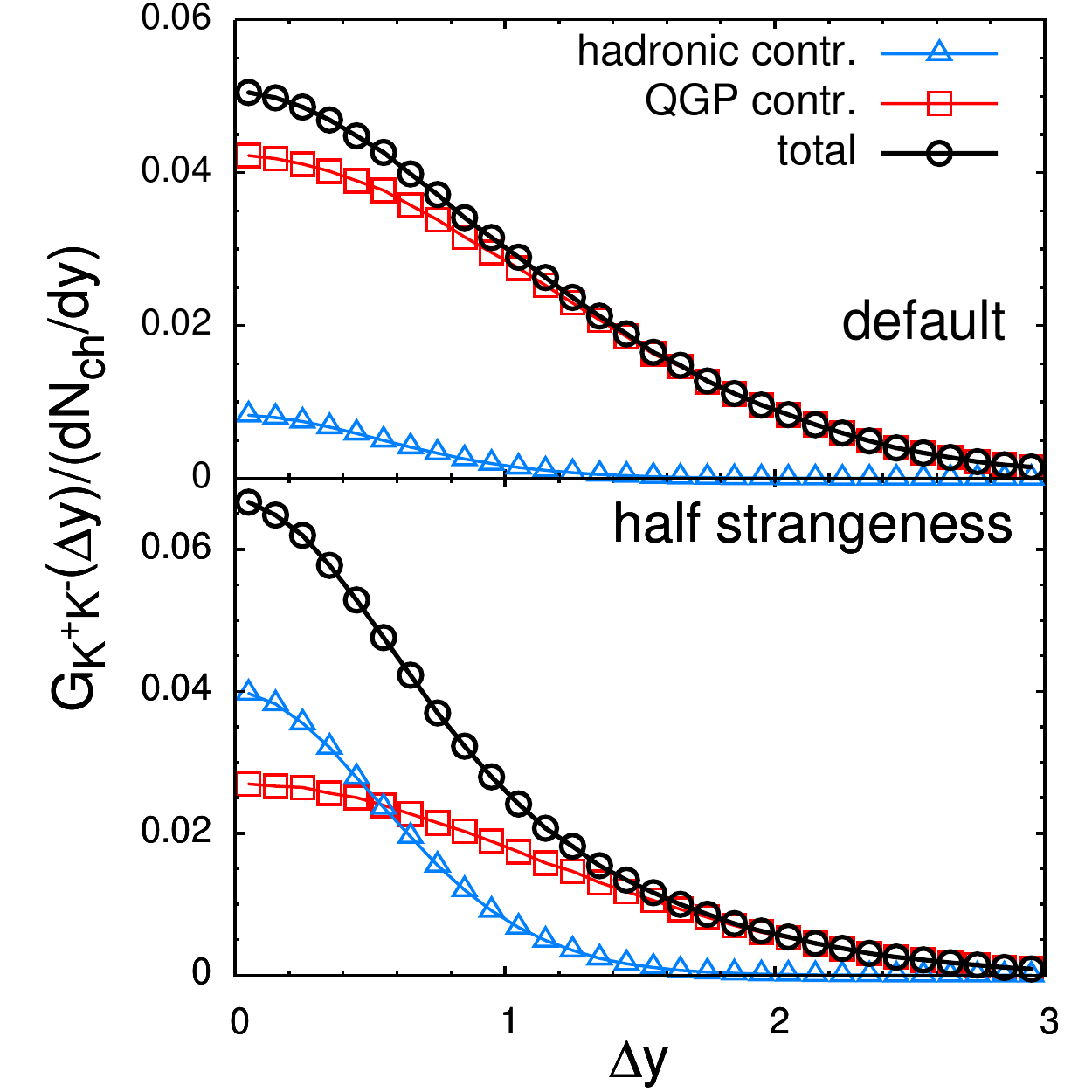}}
\caption{\label{fig:KK}
Since few strange quarks are produced during hadronization, the hadronization component for $K^+K^-$ correlations is small as shown in the upper panel. If the QGP had half the strangeness content, the hadronization component would be larger and one would have a much narrower correlation as seen in the lower panel.
}
\end{figure}

To investigate the sensitivity of the correlations to specific elements of $\chi^{\rm(QGP)}_{ab}$, several correlations are plotted in Fig.s \ref{fig:ppbar}, \ref{fig:pKminus} and \ref{fig:KK}. If the diagonal elements are reduced by a factor of two, the dip in the $p{\bar p}$ correlation disappears as seen in the middle panel of Fig. \ref{fig:ppbar}. This would be the case if the QGP had half the expected number of charge carriers. If the $ss$ component of $\chi^{\rm(QGP)}$ were halved while leaving the $uu$ and $dd$ components unchanged, the $K^+K^-$ correlation, shown in Fig. \ref{fig:KK}, becomes much narrower. This is expected, because in default case the hadronization contribution is small, and the correlation is dominated by the longer-range QGP contribution. This derives from the net number of strange quarks in the QGP being close to what one expects from an equilibrated hadron gas. If the QGP contribution is reduced, the hadronization contribution grows and results in a much narrower correlation. 

The off-diagonal elements of $\chi^{\rm(QGP)}_{ab}$ also affect final-state correlations. As an extreme case, we consider a case where the QGP is dominated by correlated quark-antiquark pairs. In that case there are large off-diagonal elements, and if the quarks are equally well paired with quarks of all species, one would expect (assuming isospin symmetry between up and down):
\begin{eqnarray}
\label{eq:offdiag}
\chi^{\rm(QGP)}_{us}&=&\chi^{\rm(QGP)}_{ds}\approx-(1/2)\chi^{\rm(QGP)}_{ss},\\
\nonumber
\chi^{\rm(QGP)}_{ud}&\approx&-\chi^{\rm(QGP)}_{uu}+(1/2)\chi_{ss}^{\rm(QGP)}.
\end{eqnarray}
For this case of strong off-diagonal elements, both $p\bar{p}$ correlations (lower panel of Fig. \ref{fig:ppbar}) and $pK^-$ correlations (lower panel of Fig. \ref{fig:pKminus}) are substantially affected. The effect comes from the fact that observing a charge does not imply observing a unit of baryon charge since all quarks are matched by antiquarks. Thus, one doesn't expect to find a balancing charge far away and the QGP components to the correlations featuring at least one baryon nearly vanish. 

The correlations $G_{\alpha\beta}(y)$ provide the means to connect $\chi_{ab}$, measured on the lattice, to observables in heavy ion collisions. The four unknown elements of $\chi^{\rm(QGP)}_{ab}$ (using isospin symmetry) along with the unknown width, $\sigma_{\rm QGP}$, can be extracted from analysis of $G_{\alpha\beta}$, which for the set of protons, kaons and pions, encompass six independent distributions, some with non-trivial shapes. The simple two-wave picture of charge production used here is only approximate and assumes that $\chi_{ab}$ stays constant during the QGP, but the ideas can be incorporated into a model with continuous evolution of $\chi$. Although the calculations presented here can only be applied at high energies, where there are roughly equal numbers of particles and antiparticles, the methods can be extended to incorporate non-zero net charge. The methods could then address the evolution of the correlations through a beam energy scan that covers the critical region for QCD.

\acknowledgements{This work was supported by the U.S. Department of Energy, Grant No. DE-FG02-03ER41259.}


\begin{thebibliography}{99}

\bibitem{rhicwhitepapers}
  I.~Arsene {\em et al.}, 
  Nucl.\ Phys.\ A {\bf 757}, 1 (2005);
  B.~B.~Back {\em et al.},
  {\em ibid.} {\bf 757}, 28 (2005);
  J.~Adams {\em et al.},
  {\em ibid.} {\bf 757}, 102 (2005);
  K.~Adcox {\em et al.},
  {\em ibid.} {\bf 757}, 184 (2005);


\bibitem{Borsanyi:2011sw}
  S.~Borsanyi, Z.~Fodor, S.~D.~Katz, S.~Krieg, C.~Ratti and K.~Szabo,
  JHEP {\bf 1201}, 138 (2012).


\bibitem{Bazavov:2012jq} 
  A.~Bazavov {\it et al.}  [HotQCD Collaboration],
  arXiv:1203.0784 [hep-lat].
  

\bibitem{Pratt:2011bc} 
  S.~Pratt,
  Phys.\ Rev.\ C {\bf 85}, 014904 (2012)
  [arXiv:1109.3647 [nucl-th]].

\bibitem{Bass:2000az}
  S.~A.~Bass, P.~Danielewicz, S.~Pratt,
  Phys.\ Rev.\ Lett.\  {\bf 85}, 2689-2692 (2000).
  [nucl-th/0005044].  

\bibitem{Cheng:2004zy} 
  S.~Cheng, S.~Petriconi, S.~Pratt, M.~Skoby, C.~Gale, S.~Jeon, V.T.~Pop and Q.~-H.~Zhang,
  Phys.\ Rev.\ C {\bf 69}, 054906 (2004)
  [nucl-th/0401008].

\bibitem{Schlichting:2010qia}
  S.~Schlichting, S.~Pratt,
  Phys.\ Rev.\  {\bf C83}, 014913 (2011).
  [arXiv:1009.4283 [nucl-th]].
  
 \bibitem{Bozek:2004dt}
  P.~Bozek,
  Phys.\ Lett.\  {\bf B609}, 247-251 (2005).
  [nucl-th/0412076].

\bibitem{Aggarwal:2010ya}
  M.~M.~Aggarwal {\it et al.} [ STAR Collaboration ],
  Phys.\ Rev.\  {\bf C82}, 024905 (2010).
  [arXiv:1005.2307 [nucl-ex]].
  
\bibitem{Adams:2003kg}
  J.~Adams {\it et al.} [ STAR Collaboration ],
  Phys.\ Rev.\ Lett.\  {\bf 90}, 172301 (2003).
  [nucl-ex/0301014].

\bibitem{Westfall:2004cq}
  G.~D.~Westfall [ STAR Collaboration ],
  J.\ Phys.\ G {\bf G30}, S345-S349 (2004).

\bibitem{Adams:2003xp}
  J.~Adams {\it et al.} [ STAR Collaboration ],
  Phys.\ Rev.\ Lett.\  {\bf 92}, 112301 (2004).
  [nucl-ex/0310004].
  
\bibitem{Adler:2003cb}
  S.~S.~Adler {\it et al.} [ PHENIX Collaboration ],
  Phys.\ Rev.\  {\bf C69}, 034909 (2004).
  [nucl-ex/0307022].
  
\end{thebibliography}
\end{document}